\documentclass[aps,prl,twocolumn,showpacs,footinbib]{revtex4-1}
\usepackage{amssymb}
\usepackage{amsmath}
\usepackage[colorlinks=true,linkcolor=blue,citecolor=blue, urlcolor=blue]{hyperref} 
\usepackage{tikz}
\usepackage{soul}
\usepackage{graphicx}
\usepackage[utf8]{inputenc}
\definecolor{uibred}{RGB}{167, 38, 47}

\newcommand{\mysection}[1]{{\vspace{10 pt}\noindent \emph{{#1}.--}}}

\begin{document}

	\title{Asymptotic freedom in a strongly interacting scalar quantum field theory in four Euclidean dimensions} 
	 
	\author{J{\"u}rgen Berges}
	\affiliation{Institut f\"{u}r Theoretische Physik, Universit\"{a}t Heidelberg, 69120 Heidelberg, Germany}
	
	\author{Razvan Gurau} 
	\affiliation{Institut f\"{u}r Theoretische Physik, Universit\"{a}t Heidelberg, 69120 Heidelberg, Germany}
	
	\author{Thimo Preis} \email{preis@thphys.uni-heidelberg.de}
	\affiliation{Institut f\"{u}r Theoretische Physik, Universit\"{a}t Heidelberg, 69120 Heidelberg, Germany}

\begin{abstract}
We show that scalar quantum field theory in four Euclidean dimensions with global $O(N)^3$ symmetry and imaginary tetrahedral coupling is asymptotically free and bounded from below in the large-$N$ limit. 
While the Hamiltonian is non-Hermitian, the full quantum effective action for the large-$N$ theory only depends on the square of that coupling which is real. A perturbative analysis uncovers that the renormalization group flow of the quartic couplings connects a Gaussian ultraviolet fixed point to a strongly interacting theory in the infrared. This realizes a renormalizable field theory which exhibits non-trivial dynamics, such as direct scattering, while still being analytically tractable also non-perturbatively. 
Our findings open up a way to address outstanding problems in strongly coupled theories from first principles.
\end{abstract}
	
\maketitle

\mysection{Introduction} 
Asymptotic freedom in the theory of the strong interaction is a hallmark for the fundamental description of nature encoded in the Standard Model of particle physics~\cite{Gross:1973id,Politzer:1973fx}. 
The phenomenon describes the change of the interaction strength with the characteristic energy scale, such that the theory becomes non-interacting at high energy-momenta. In turn, interactions become strong at low energies with striking phenomenological implications such as the confinement of hadronic matter~\cite{Wilson:1974sk}. While dynamical strong-interaction phenomena, like quantum bounds on crucial observables such as viscosities~\cite{Policastro:2001yc} or relaxation times~\cite{Sekino:2008he}, have been suggested by holographic models, their analysis based directly on quantum theories with scale-dependent interactions still remains elusive.  

Asymptotic freedom is linked to local space-time symmetries of non-Abelian gauge fields in the Standard Model. However, it can in principle be realized even for simple scalar field theories in four space-time dimensions if the condition of Hermiticity is relaxed~\cite{Coleman:1973sx}. A known example is scalar field theory with negative quartic self-interaction~\cite{Symanzik:1973hx,Romatschke:2022jqg}. However, the energy of the system cannot be bounded from below and the theory is considered to be ill-defined~\cite{Coleman:1973sx}. Nevertheless, it has been suggested that the parity and time-reversal symmetric theory may still posses a real and positive Eigenspectrum, obtained by an analytic continuation procedure from the corresponding bounded theory with positive interaction~\cite{Bender:1998ke,Ai:2022csx,Romatschke:2022jqg,Bender:2018pbv}. This fosters the discussion about the remarkable possibility of a larger class of well-defined asymptotically free theories without relying on Hermiticity, which is a sufficient but not necessary condition. The notion of asymptotically free scalar field theories may also provide an important ingredient in our understanding of the scalar Higgs particle with potential consequences also for physics beyond the Standard Model~\cite{Mavromatos:2020bbq,Alexandre:2020gah,Fring:2021zci}.

\begin{figure}[t!]
\includegraphics[width=0.9\columnwidth]{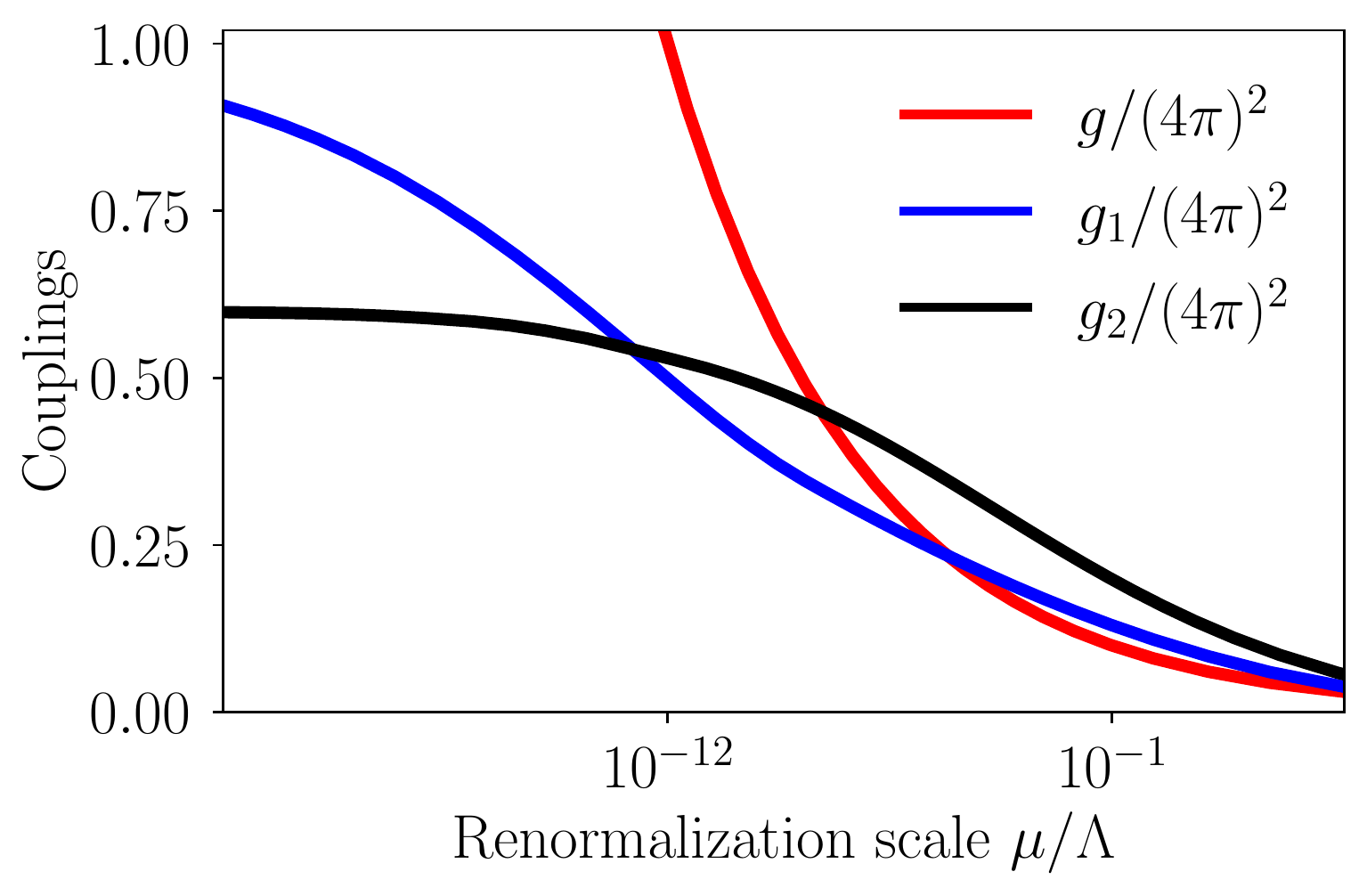}
\caption{Renormalization group flow of the couplings as a function of the renormalization scale $\mu/\Lambda$. The flow connects the asymptotically free UV with the strongly interacting IR, corresponding to the upper (red) separatrices of Fig.~\ref{fig.on3_betaflow_gcomplex}. 
}
\label{fig.UVfreedom}
\end{figure}

In this work, we show that both asymptotic freedom and boundedness from below can be realized in scalar quantum field theory in four Euclidean dimensions. We demonstrate our finding for a theory with global $O(N)^3$ symmetry, where the dynamics can be studied via a systematic large-$N$ limit in the number of field components $\varphi_{\mathbf{a}=(a^1, a^2 ,a^3)}$ with $a^{i=1,2,3}=1,\dots,N$. To uncover its scale dependence, we analyze the renormalization group flow for the tensor field which features two real quartic couplings, $g_1$ and $g_2$ (``pillow/double-trace"), and an imaginary coupling, $ig$ (``tetrahedral"), without the need for any analytic continuation. It turns out that all the couplings exhibit asymptotic freedom in a regime governed by the flow of the imaginary coupling, leading to a renormalizable field theory with a well-defined ultraviolet (UV) limit and a strongly coupled infrared (IR). The renormalization group flow of the three couplings in this case as a function of the renormalization scale $\mu$ with UV scale $\Lambda$ is displayed in Fig.~\ref{fig.UVfreedom}.

In contrast to the more conventional $N$-component vector field theories~\cite{Moshe:2003xn}, such as describing the Higgs sector of the Standard Model for finite $N=4$, the real bosonic tensor field $\varphi_{\mathbf{a}}$ exhibits non-trivial dynamics involving direct scattering already at leading order in the large-$N$ expansion, while still being analytically tractable~\cite{Bonzom:2011zz,Gurau:2011xp}. The latter represents also a significant advantage compared to the large-$N$ limit in matrix models involving all planar Feynman diagrams for quantum corrections~\cite{tHooft:1973alw,Brezin:1977sv}. It has been remarked~\cite{Witten:2016iux} that tensor quantum mechanical models provide an alternative to the Sachdev-Ye-Kitaev model~\cite{Sachdev:1992fk,Kitaev} without disorder. Subsequently tensor field theories have been shown to display strongly coupled infrared fixed points in lower Euclidean space-time dimensions $d<4$~\cite{Klebanov:2016xxf,Giombi:2017dtl,Benedetti:2019eyl,Benedetti:2021wzt}. In the following, we reveal that this theory becomes highly non-trivial in $d=4$. 

\mysection{Tensor field theory with imaginary tetrahedral coupling}
The scalar field $\varphi_{\mathbf{a}}(x)$, with $x$ a Euclidean space-time point, is a real tensor of rank three transforming in the tri-fundamental representation of $O(N)^3$~\cite{Giombi:2017dtl}. 
The classical action of the theory reads in four dimensions:
\begin{align}
    S[\varphi] &=  \int d^4 x \bigg\{ \frac{1}{2} \varphi_{\mathbf{a}}(x)(-\partial^2 +\bar{m}^2) \varphi_{\mathbf{a}}(x) \nonumber\\
    &+\frac{1}{4}\left(\bar{g}_1 \hat{P}^{(1)}_{\mathbf{a} \mathbf{b};\mathbf{c} \mathbf{d} } + \bar{g}_2  \hat{P}^{(2)}_{\mathbf{a} \mathbf{b};\mathbf{c} \mathbf{d} }+i\bar{g} \hat{\delta}^t_{\mathbf{a} \mathbf{b}\mathbf{c} \mathbf{d} }\right)\nonumber\\
    &\times \;\varphi_{\mathbf{a}}(x) \varphi_{\mathbf{b}}(x) \varphi_{\mathbf{c}}(x) \varphi_{\mathbf{d}}(x) \bigg\}\;,
    \label{eq:S}
\end{align}
where $\bar{m}$ denotes the real mass parameter, and we take the coupling parameters $\bar{g}_1$, $\bar{g}_2$ and $\bar{g}$ to be real such that $i \bar{g}$ appearing in Eq.~(\ref{eq:S}) is purely imaginary. The corresponding three interaction terms in the action reflect the three $O(N)^3$ invariant contraction patterns (pillow, double-trace and tetrahedral):
\begin{align}
    \hat{\delta}^p_{\mathbf{a} \mathbf{b}; \mathbf{c}\mathbf{d}}&= \frac{1}{3N^2} \sum_{i=1}^3 \delta_{a^i c^i} \delta_{b^i d^i} \prod_{j\neq i} \delta_{a^jb^j} \delta_{c^jd^j} \; ,\nonumber \\
    \hat{\delta}^d_{\mathbf{a}\mathbf{b};\mathbf{c} \mathbf{d}} &=N^{-3} \prod_{i=1}^3 \delta_{a^ib^i} \prod_{j=1}^3 \delta_{c^jd^j} \; ,  \\
    \hat{\delta}^t_{\mathbf{a} \mathbf{b} \mathbf{c} \mathbf{d}} &= N^{-3/2} \delta_{a^1 b^1} \delta_{c^1 d^1} \delta_{a^2 c^2} \delta_{b^2d^2} \delta_{a^3 d^3} \delta_{b^3 c^3} \; , \nonumber
\end{align}
which relate to the orthonormal projectors  $\hat{P}^{(1)}=3(\hat{\delta}^p-\hat{\delta}^d)$ and $\hat{P}^{(2)}=\hat{\delta}^d$.
Their scaling with $N$ is determined such that the theory has a well-defined large-$N$ limit.

The full quantum theory is described in path-integral quantization by a functional integral whose norm,
\begin{align}
    \left|\int \mathcal{D}\varphi\, e^{-S[\varphi;\bar{g}_1,\bar{g}_2,i\bar{g}]}\right| \,
    &\leq \int \mathcal{D}\varphi\, \left|e^{-S[\varphi;\bar{g}_1,\bar{g}_2,i\bar{g}]}\right|\nonumber\\
    &=\int \mathcal{D} \varphi\, e^{- S[\varphi;\bar{g}_1,\bar{g}_2, i\bar{g} \equiv 0]} \;,
\end{align}
is bounded by the corresponding model with real action \mbox{$S[\varphi;\bar{g}_1,\bar{g}_2, i\bar{g} \equiv 0]$}, since the couplings $\bar{g}_1$ and $\bar{g}_2$ are real. The theory is thus stable and bounded from below if $\bar{g}_1$ and $\bar{g}_2$ are positive. 

The imaginary tetrahedral coupling parameter $i\bar{g}$ introduces an oscillatory term in the functional integral. This is difficult to compute with standard importance sampling techniques, where the situation is reminiscent of the notorious sign problem, as e.g.\ in quantum chromodynamics at non-zero baryon number density~\cite{deForcrand:2009zkb}. However, in our case it poses no particular problem for first-principles computations using large-$N$ techniques.     

The quantum theory in the large-$N$ limit is conveniently described in terms of a free-energy functional $\Gamma[G]$, which is the two-particle irreducible ($2$PI) effective action  \cite{Cornwall:1974vz,Benedetti:2018goh}. It is a functional of the full propagator $G_{\mathbf{a} \mathbf{b}}(x,y)$, which is determined self-consistently from a variational principle $\delta \Gamma[G]/\delta G_{\mathbf{a} \mathbf{b}}(x,y) =0$. The $2$PI effective action can be written as:
\begin{equation}
    \Gamma[G] = \frac{1}{2} \mathrm{Tr} \ln G^{-1} + \frac{1}{2} \mathrm{Tr}[G^{-1}_{0} G] + \Gamma_2[G] \;.
\end{equation}
Here the classical inverse propagator is given by $G^{-1}_{0,\mathbf{a}\mathbf{b}}(x,y) \equiv \delta^2 S/\delta \varphi_{\mathbf{a}}(x) \delta \varphi_{\mathbf{b}}(y)$, and $\Gamma_2[G]$ contains all $2$PI contributions (Feynman graphs that are not disconnected by cutting two propagator lines).
Considering the $O(N)^3$ symmetric regime with $G_{\mathbf{a}\mathbf{b}}(x,y) =G(x,y) \prod_{i=1}^3 \delta_{a^i b^i} $, we get at leading order large-$N$:
\begin{align}
    \frac{\Gamma_2[G]}{N^{3}} &= \frac{\bar{g}_2}{4}\int d^4 x \, G^2(x,x) \nonumber\\
    &+\frac{\bar{g}^2}{8} \int d^4x d^4 y \, G^4(x,y)\; .
\label{eq:Gamma2}
\end{align}
The self-consistent propagator reads in Fourier space:
\begin{align}
    G^{-1}(p) &=p^2+\bar{m}^2+\bar{g}_2 \int \frac{d^4 q}{(2\pi)^4}\, G(q) \nonumber\\
    &+\bar{g}^2 \int \frac{d^4 q}{(2\pi)^4}\frac{d^4 k}{(2\pi)^4}\,  G(q) G(k) G(p+q+k)\label{eq.sdeq_LO} \;.
\end{align}
Apart from the different factors and couplings in front of the tadpole term $\sim \bar{g}_2$ and sunset (or melon) contribution $\sim \bar{g}^2$, this equation is of the same form as the one obtained in the $O(N)$ symmetric $N$-vector model from a $2$PI two-loop expansion~\cite{Berges:2004yj}. We emphasize that, while the corresponding equation is only valid at weak coupling in the $N$-vector model, Eq.~(\ref{eq.sdeq_LO}) is exact in the large-$N$ limit of the $O(N)^3$ tensor field theory we consider. 

In the large-$N$ limit, but at all orders in perturbation theory, the tetrahedral coupling enters the $2$PI effective action (\ref{eq:Gamma2}), and consequentially the two- and four-point correlation functions, only quadratically (cf.\ also the discussion in Ref.~\cite{Benedetti:2019eyl}). Therefore, the purely imaginary tetrahedral coupling simply leads in the large-$N$ quantum effective action to some sign changes with respect to the real case. 

\begin{figure*}[t!]
\centering
\includegraphics[width=0.95\linewidth]{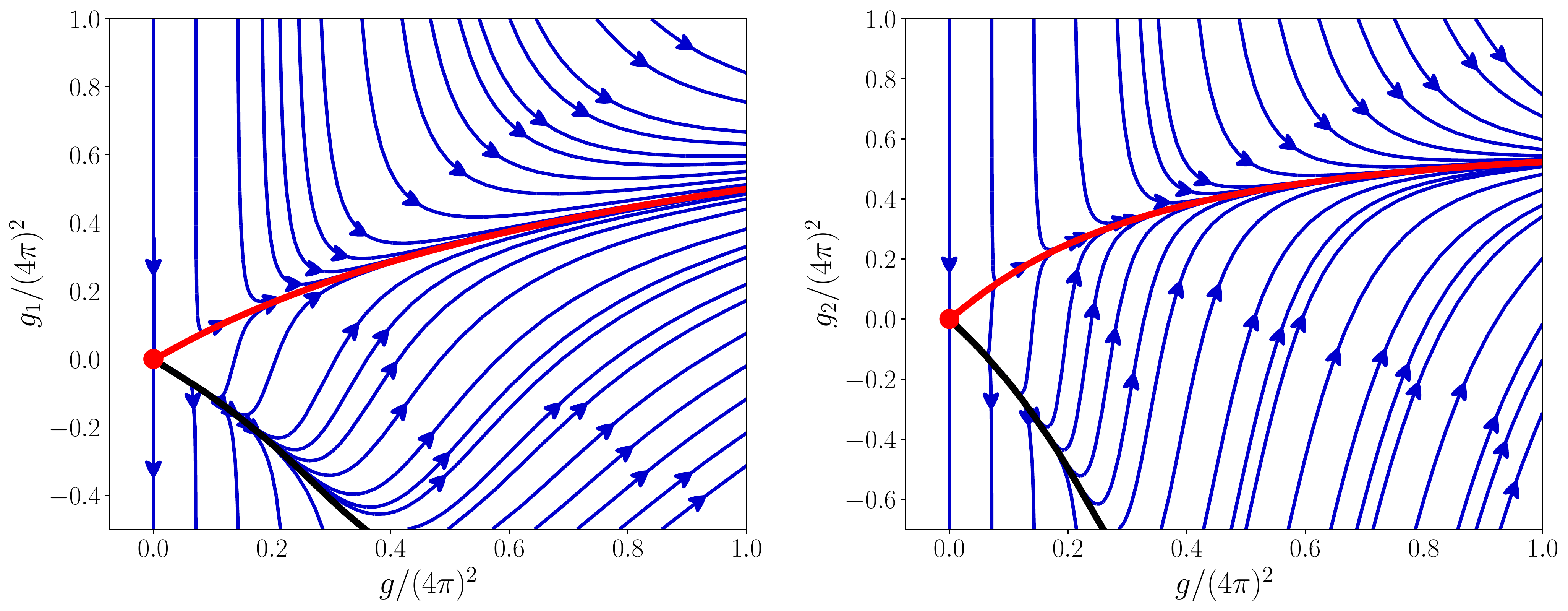}
\caption{Flow trajectories of $g_1$ (left) and $g_2$ (right) as obtained from Eq. (\ref{eq.betas_kappa}), where arrows point in the direction of smaller scale $\mu$. Bold upper (red) and lower (black) lines denote separatrices, such that all models with coupling values at and below the red separatrix exhibit asymptotic freedom, while for values above the red line a Landau pole is realized in the UV. All non-zero couplings above the black separatrix feature a strongly interacting IR with positive values.
}
\label{fig.on3_betaflow_gcomplex}
\end{figure*} 
The next-to-leading order contribution to the 2PI effective action is suppressed by $N^{-1/2}$~\cite{Benedetti:2018goh}. It consists of double tadpoles with only one tetrahedral vertex
such that the next-to-leading contribution  exhibits an imaginary term. Its presence leads to two classes of subleading imaginary contributions in correlation functions. First, one gets a shift in the bare mass. However, since the physical mass is fixed by a real infrared renormalization condition, in our case this is without consequences. Second, one obtains a subleading term in the four-point kernel corresponding to the tetrahedral pattern of identification of tensor indices. The latter backreacts on the two-loop $\beta$-functions (cf.\ Eq.\ (2.40) of Ref.~\cite{Benedetti:2020sye}), and brings in corrections $\sim i N^{-1/2}$.
While imaginary contributions arise, the large-$N$ behavior is not singular and the corrections with respect to the leading-order behavior are organized in powers of $N^{-1/2}$, such that the Euclidean model remains asymptotically free and stable: Although they acquire small imaginary parts, the couplings $g_1,g_2$ maintain a positive real part and the $N^{-1/2}$ corrections do not spoil the boundedness from below of the real part of the action. Therefore, the real effective action (\ref{eq:Gamma2}) is exact in the large-$N$ limit which is well-controlled.

 \mysection{Renormalization group flow}
 To demonstrate that the theory realizes asymptotic freedom, we study the perturbative renormalization group flow of the quartic couplings. Their flow is described in terms of $\beta$-functions, which encode how quantum corrections change the corresponding renormalized couplings $g, g_1,$ and $g_2$ with the renormalization scale $\mu$: $\beta\equiv\mu \partial_\mu g(\mu)$ and analogously for $g_1(\mu)$ and $g_2(\mu)$.
 The corresponding system of $\beta$ functions of the model has been investigated in $d<4$, see Refs.~\cite{Giombi:2017dtl,Benedetti:2019eyl} and subsequent work. 

If one considers all couplings to be real, in $d = 4$ the theory exhibits triviality, i.e.~it becomes non-interacting corresponding to a Gaussian fixed point in the IR as \mbox{$\mu \rightarrow 0$}. For the case of an imaginary tetrahedral coupling we are discussing in this work, we find a strikingly different behavior. We obtain the corresponding beta functions in the large-$N$ limit and at two loop order for the massless theory, which is sufficient for our considerations, from Ref.~\cite{Giombi:2017dtl} as: 
\begin{subequations} 
\label{eq.betas_kappa}
\begin{align}
    \beta &= -\frac{2g^3}{(4\pi)^4}  \label{eq:betag}  \;, \\
    \beta_1 &=2\left(\frac{ g^2_1-g^2 }{(4\pi)^2} +\frac{g_1 g^2 }{(4\pi)^4}\right) 
    \; , \\
   \beta_2 &= 2\left( \frac{g^2_2-3g^2}{(4\pi)^2} + \frac{5 g_2 g^2}{(4\pi)^4}\right)   \;.
\end{align}
\end{subequations}
Since the $\beta$-function for the tetrahedral coupling (\ref{eq:betag}) is independent of the other couplings at this order, for given coupling value $g_\Lambda$ at some UV scale $\Lambda$ its flow is solved by:
\begin{equation}
    g^2(\mu) = \frac{g_\Lambda^2}{1-\frac{4g_\Lambda^2}{(4\pi)^4} \ln\left(\Lambda/\mu\right)} \;.
\end{equation}
This demonstrates that $g^2(\mu)$ approaches zero logarithmically as $\mu$ increases. Conversely, $g^2(\mu)$ grows towards the IR and, to lowest order in perturbation theory, the coupling diverges at $\mu_*=\Lambda \exp[-(4\pi)^4/(4g_\Lambda^2)]$. 
At two-loop order in the large-$N$ limit we additionally infer that, corresponding to the negative sign of the $\beta$-function for the tetrahedral coupling, the anomalous dimension~\cite{Giombi:2017dtl} $\gamma_\phi = -g^2/(2(4\pi)^4)$ is also negative, similar to other theories with negative $\beta$-functions~\cite{Christiansen:2017cxa,Zinn-Justin:1989rgp}. 

The flow of the other two couplings, which depend on the values of $g$, is depicted in 
Fig.~\ref{fig.on3_betaflow_gcomplex}. The blue lines represent the values of $g_1(\mu)/(4 \pi)^2$ (left graph) and $g_2(\mu)/(4 \pi)^2$ (right graph) for corresponding values of $g(\mu)/(4 \pi)^2$, where arrows point towards smaller (IR) values of $\mu$.
The (upper) red and (lower) black lines denote separatrices dividing the flow diagrams into distinct regions. For all $g > 0$ above the black separatrix, the theory exhibits a non-trivial IR behavior for $g_1$ and $g_2$, with vanishing $\beta$-functions for the fixed values $g_1/(4\pi)^2\to 1$ and $g_2/(4\pi)^2\to 3/5$. Importantly, in the sector comprised between the red and the black separatrices, the renormalization group flow connects an ultraviolet attractive Gaussian fixed point (red dot) to the non-trivial infrared theory. All quartic couplings $g, g_1,$ and $g_2$ are thus found to exhibit asymptotic freedom along these trajectories, which a posteriori justifies the validity of our two loop analysis of the corresponding $\beta$ functions in this regime. In particular, along the red separatrix the couplings are positive at all scales. This realizes a renormalizable scalar field theory, where the non-tetrahedral quartic couplings are real and positive, thereby ensuring that the corresponding path integral is bounded from below, with a well-defined UV limit and a strongly coupled IR. 
The corresponding flow of couplings as a function of the renormalization scale $\mu$ is given in Fig.~\ref{fig.UVfreedom}. 
	
\mysection{Conclusion and outlook}
Our findings establish asymptotic freedom in a well-defined strongly interacting scalar quantum field theory in four Euclidean dimensions. This is a remarkable result also in view of the fact that the tetrahedral coupling driving the renormalization group flow in this regime is purely imaginary. Despite this, the full quantum effective action for the theory in the large-$N$ limit only depends on the square of that coupling which is real. Correspondingly, the renormalization group $\beta$-functions we discussed are also real. These findings support the exciting possibility of an extended class of well-defined theories where the standard condition of Hermiticity is relaxed \cite{Fisher:1978pf,Gromov:2018hut}.  

The large-$N$ limit of the tensor field theory we considered is so interesting because it is already highly non-trivial, involving crucial processes such as direct scatterings for the dynamics, and analytically tractable also non-perturbatively. This is in contrast to the large-$N$ limit in $N$-vector models, where collisional processes are absent; or in matrix models, where no general closed form for the non-trivial leading-order dynamics is known. 

So far, our findings concern the Euclidean field theory, which is relevant for vacuum or equilibrium properties. It is an important open question to what extent our results can be taken over to Minkowski space-time, in particular, since our arguments on boundedness use well-defined properties of the Euclidean theory. This would make our theory a versatile theoretical laboratory to investigate crucial open dynamical questions, such as real-time dynamics in the strong-coupling regime. While holographic models can give very important insights at strong coupling~\cite{Casalderrey-Solana:2011dxg,Berges:2020fwq}, it is typically difficult to include the relevant changes of the interaction strength with the scale as is the case in an asymptotically free theory. 

At lowest non-trivial order in perturbation theory, we found that the tetrahedral coupling grows without bound in the IR. An important question is whether the renormalization group flow is altered when the coupling becomes non-perturbatively large. 
This question can be answered from the full large-$N$ result, which resums contributions from perturbation theory to all orders in the coupling. In order to derive the exact  $\beta$-functions in the large-$N$ limit, one needs to deal appropriately with the wave function renormalization in conjunction with Eq.~\eqref{eq.sdeq_LO}. Once this is done, the all-order running of the four-point couplings should be accessible by techniques similar to the ones employed in Ref.~\cite{Benedetti:2019eyl}. 
A striking possible outcome would be that the tetrahedral coupling $g$ approaches a finite IR value in the non-perturbative regime. Together with our findings for $g_1$ and $g_2$ this would correspond to a proper infrared fixed point. Its universal properties then allow one to make very powerful statements about all possible strong-coupling theories in the same universality class, which only depends on general properties such as symmetries. 

\mysection{Acknowledgments} 
This work is supported by the Deutsche Forschungsgemeinschaft (DFG, German Research Foundation) under Germany’s Excellence
Strategy EXC2181/1-390900948 (the Heidelberg STRUCTURES Excellence Cluster), and the Collaborative Research Center,
Project-ID 273811115, SFB 1225 ISOQUANT, as well as the European Research Council (ERC) under the European
Union’s Horizon 2020 research and innovation program (grant agreement No818066).

\bibliography{master.bib}

\begin{thebibliography}{37}%
\makeatletter
\providecommand \@ifxundefined [1]{%
 \@ifx{#1\undefined}
}%
\providecommand \@ifnum [1]{%
 \ifnum #1\expandafter \@firstoftwo
 \else \expandafter \@secondoftwo
 \fi
}%
\providecommand \@ifx [1]{%
 \ifx #1\expandafter \@firstoftwo
 \else \expandafter \@secondoftwo
 \fi
}%
\providecommand \natexlab [1]{#1}%
\providecommand \enquote  [1]{``#1''}%
\providecommand \bibnamefont  [1]{#1}%
\providecommand \bibfnamefont [1]{#1}%
\providecommand \citenamefont [1]{#1}%
\providecommand \href@noop [0]{\@secondoftwo}%
\providecommand \href [0]{\begingroup \@sanitize@url \@href}%
\providecommand \@href[1]{\@@startlink{#1}\@@href}%
\providecommand \@@href[1]{\endgroup#1\@@endlink}%
\providecommand \@sanitize@url [0]{\catcode `\\12\catcode `\$12\catcode
  `\&12\catcode `\#12\catcode `\^12\catcode `\_12\catcode `\%12\relax}%
\providecommand \@@startlink[1]{}%
\providecommand \@@endlink[0]{}%
\providecommand \url  [0]{\begingroup\@sanitize@url \@url }%
\providecommand \@url [1]{\endgroup\@href {#1}{\urlprefix }}%
\providecommand \urlprefix  [0]{URL }%
\providecommand \Eprint [0]{\href }%
\providecommand \doibase [0]{http://dx.doi.org/}%
\providecommand \selectlanguage [0]{\@gobble}%
\providecommand \bibinfo  [0]{\@secondoftwo}%
\providecommand \bibfield  [0]{\@secondoftwo}%
\providecommand \translation [1]{[#1]}%
\providecommand \BibitemOpen [0]{}%
\providecommand \bibitemStop [0]{}%
\providecommand \bibitemNoStop [0]{.\EOS\space}%
\providecommand \EOS [0]{\spacefactor3000\relax}%
\providecommand \BibitemShut  [1]{\csname bibitem#1\endcsname}%
\let\auto@bib@innerbib\@empty
\bibitem [{\citenamefont {Gross}\ and\ \citenamefont
  {Wilczek}(1973)}]{Gross:1973id}%
  \BibitemOpen
  \bibfield  {author} {\bibinfo {author} {\bibfnamefont {D.~J.}\ \bibnamefont
  {Gross}}\ and\ \bibinfo {author} {\bibfnamefont {F.}~\bibnamefont
  {Wilczek}},\ }\href {\doibase 10.1103/PhysRevLett.30.1343} {\bibfield
  {journal} {\bibinfo  {journal} {Phys. Rev. Lett.}\ }\textbf {\bibinfo
  {volume} {30}},\ \bibinfo {pages} {1343} (\bibinfo {year}
  {1973})}\BibitemShut {NoStop}%
\bibitem [{\citenamefont {Politzer}(1973)}]{Politzer:1973fx}%
  \BibitemOpen
  \bibfield  {author} {\bibinfo {author} {\bibfnamefont {H.~D.}\ \bibnamefont
  {Politzer}},\ }\href {\doibase 10.1103/PhysRevLett.30.1346} {\bibfield
  {journal} {\bibinfo  {journal} {Phys. Rev. Lett.}\ }\textbf {\bibinfo
  {volume} {30}},\ \bibinfo {pages} {1346} (\bibinfo {year}
  {1973})}\BibitemShut {NoStop}%
\bibitem [{\citenamefont {Wilson}(1974)}]{Wilson:1974sk}%
  \BibitemOpen
  \bibfield  {author} {\bibinfo {author} {\bibfnamefont {K.~G.}\ \bibnamefont
  {Wilson}},\ }\href {\doibase 10.1103/PhysRevD.10.2445} {\bibfield  {journal}
  {\bibinfo  {journal} {Phys. Rev. D}\ }\textbf {\bibinfo {volume} {10}},\
  \bibinfo {pages} {2445} (\bibinfo {year} {1974})}\BibitemShut {NoStop}%
\bibitem [{\citenamefont {Policastro}\ \emph {et~al.}(2001)\citenamefont
  {Policastro}, \citenamefont {Son},\ and\ \citenamefont
  {Starinets}}]{Policastro:2001yc}%
  \BibitemOpen
  \bibfield  {author} {\bibinfo {author} {\bibfnamefont {G.}~\bibnamefont
  {Policastro}}, \bibinfo {author} {\bibfnamefont {D.~T.}\ \bibnamefont {Son}},
  \ and\ \bibinfo {author} {\bibfnamefont {A.~O.}\ \bibnamefont {Starinets}},\
  }\href {\doibase 10.1103/PhysRevLett.87.081601} {\bibfield  {journal}
  {\bibinfo  {journal} {Phys. Rev. Lett.}\ }\textbf {\bibinfo {volume} {87}},\
  \bibinfo {pages} {081601} (\bibinfo {year} {2001})},\ \Eprint
  {http://arxiv.org/abs/hep-th/0104066} {arXiv:hep-th/0104066} \BibitemShut
  {NoStop}%
\bibitem [{\citenamefont {Sekino}\ and\ \citenamefont
  {Susskind}(2008)}]{Sekino:2008he}%
  \BibitemOpen
  \bibfield  {author} {\bibinfo {author} {\bibfnamefont {Y.}~\bibnamefont
  {Sekino}}\ and\ \bibinfo {author} {\bibfnamefont {L.}~\bibnamefont
  {Susskind}},\ }\href {\doibase 10.1088/1126-6708/2008/10/065} {\bibfield
  {journal} {\bibinfo  {journal} {JHEP}\ }\textbf {\bibinfo {volume} {10}},\
  \bibinfo {pages} {065} (\bibinfo {year} {2008})},\ \Eprint
  {http://arxiv.org/abs/0808.2096} {arXiv:0808.2096 [hep-th]} \BibitemShut
  {NoStop}%
\bibitem [{\citenamefont {Coleman}\ and\ \citenamefont
  {Gross}(1973)}]{Coleman:1973sx}%
  \BibitemOpen
  \bibfield  {author} {\bibinfo {author} {\bibfnamefont {S.~R.}\ \bibnamefont
  {Coleman}}\ and\ \bibinfo {author} {\bibfnamefont {D.~J.}\ \bibnamefont
  {Gross}},\ }\href {\doibase 10.1103/PhysRevLett.31.851} {\bibfield  {journal}
  {\bibinfo  {journal} {Phys. Rev. Lett.}\ }\textbf {\bibinfo {volume} {31}},\
  \bibinfo {pages} {851} (\bibinfo {year} {1973})}\BibitemShut {NoStop}%
\bibitem [{\citenamefont {Symanzik}(1973)}]{Symanzik:1973hx}%
  \BibitemOpen
  \bibfield  {author} {\bibinfo {author} {\bibfnamefont {K.}~\bibnamefont
  {Symanzik}},\ }\href {\doibase 10.1007/BF02788323} {\bibfield  {journal}
  {\bibinfo  {journal} {Lett. Nuovo Cim.}\ }\textbf {\bibinfo {volume} {6S2}},\
  \bibinfo {pages} {77} (\bibinfo {year} {1973})}\BibitemShut {NoStop}%
\bibitem [{\citenamefont {Romatschke}(2022)}]{Romatschke:2022jqg}%
  \BibitemOpen
  \bibfield  {author} {\bibinfo {author} {\bibfnamefont {P.}~\bibnamefont
  {Romatschke}},\ }\href@noop {} {\  (\bibinfo {year} {2022})},\ \Eprint
  {http://arxiv.org/abs/2211.15683} {arXiv:2211.15683 [hep-th]} \BibitemShut
  {NoStop}%
\bibitem [{\citenamefont {Bender}\ and\ \citenamefont
  {Boettcher}(1998)}]{Bender:1998ke}%
  \BibitemOpen
  \bibfield  {author} {\bibinfo {author} {\bibfnamefont {C.~M.}\ \bibnamefont
  {Bender}}\ and\ \bibinfo {author} {\bibfnamefont {S.}~\bibnamefont
  {Boettcher}},\ }\href {\doibase 10.1103/PhysRevLett.80.5243} {\bibfield
  {journal} {\bibinfo  {journal} {Phys. Rev. Lett.}\ }\textbf {\bibinfo
  {volume} {80}},\ \bibinfo {pages} {5243} (\bibinfo {year} {1998})},\ \Eprint
  {http://arxiv.org/abs/physics/9712001} {arXiv:physics/9712001} \BibitemShut
  {NoStop}%
\bibitem [{\citenamefont {Ai}\ \emph {et~al.}(2022)\citenamefont {Ai},
  \citenamefont {Bender},\ and\ \citenamefont {Sarkar}}]{Ai:2022csx}%
  \BibitemOpen
  \bibfield  {author} {\bibinfo {author} {\bibfnamefont {W.-Y.}\ \bibnamefont
  {Ai}}, \bibinfo {author} {\bibfnamefont {C.~M.}\ \bibnamefont {Bender}}, \
  and\ \bibinfo {author} {\bibfnamefont {S.}~\bibnamefont {Sarkar}},\ }\href
  {\doibase 10.1103/PhysRevD.106.125016} {\bibfield  {journal} {\bibinfo
  {journal} {Phys. Rev. D}\ }\textbf {\bibinfo {volume} {106}},\ \bibinfo
  {pages} {125016} (\bibinfo {year} {2022})},\ \Eprint
  {http://arxiv.org/abs/2209.07897} {arXiv:2209.07897 [hep-th]} \BibitemShut
  {NoStop}%
\bibitem [{\citenamefont {Bender}\ \emph {et~al.}(2018)\citenamefont {Bender},
  \citenamefont {Hassanpour}, \citenamefont {Klevansky},\ and\ \citenamefont
  {Sarkar}}]{Bender:2018pbv}%
  \BibitemOpen
  \bibfield  {author} {\bibinfo {author} {\bibfnamefont {C.~M.}\ \bibnamefont
  {Bender}}, \bibinfo {author} {\bibfnamefont {N.}~\bibnamefont {Hassanpour}},
  \bibinfo {author} {\bibfnamefont {S.~P.}\ \bibnamefont {Klevansky}}, \ and\
  \bibinfo {author} {\bibfnamefont {S.}~\bibnamefont {Sarkar}},\ }\href
  {\doibase 10.1103/PhysRevD.98.125003} {\bibfield  {journal} {\bibinfo
  {journal} {Phys. Rev. D}\ }\textbf {\bibinfo {volume} {98}},\ \bibinfo
  {pages} {125003} (\bibinfo {year} {2018})},\ \Eprint
  {http://arxiv.org/abs/1810.12479} {arXiv:1810.12479 [hep-th]} \BibitemShut
  {NoStop}%
\bibitem [{\citenamefont {Mavromatos}(2020)}]{Mavromatos:2020bbq}%
  \BibitemOpen
  \bibfield  {author} {\bibinfo {author} {\bibfnamefont {N.~E.}\ \bibnamefont
  {Mavromatos}},\ }\href {\doibase 10.1088/1742-6596/2038/1/012019} {\bibfield
  {journal} {\bibinfo  {journal} {J. Phys. Conf. Ser.}\ }\textbf {\bibinfo
  {volume} {2038}},\ \bibinfo {pages} {012019} (\bibinfo {year} {2020})},\
  \Eprint {http://arxiv.org/abs/2010.15790} {arXiv:2010.15790 [hep-ph]}
  \BibitemShut {NoStop}%
\bibitem [{\citenamefont {Alexandre}\ \emph {et~al.}(2020)\citenamefont
  {Alexandre}, \citenamefont {Ellis},\ and\ \citenamefont
  {Millington}}]{Alexandre:2020gah}%
  \BibitemOpen
  \bibfield  {author} {\bibinfo {author} {\bibfnamefont {J.}~\bibnamefont
  {Alexandre}}, \bibinfo {author} {\bibfnamefont {J.}~\bibnamefont {Ellis}}, \
  and\ \bibinfo {author} {\bibfnamefont {P.}~\bibnamefont {Millington}},\
  }\href {\doibase 10.1103/PhysRevD.102.125030} {\bibfield  {journal} {\bibinfo
   {journal} {Phys. Rev. D}\ }\textbf {\bibinfo {volume} {102}},\ \bibinfo
  {pages} {125030} (\bibinfo {year} {2020})},\ \Eprint
  {http://arxiv.org/abs/2006.06656} {arXiv:2006.06656 [hep-th]} \BibitemShut
  {NoStop}%
\bibitem [{\citenamefont {Fring}\ and\ \citenamefont
  {Taira}(2021)}]{Fring:2021zci}%
  \BibitemOpen
  \bibfield  {author} {\bibinfo {author} {\bibfnamefont {A.}~\bibnamefont
  {Fring}}\ and\ \bibinfo {author} {\bibfnamefont {T.}~\bibnamefont {Taira}},\
  }\href {\doibase 10.1088/1742-6596/2038/1/012010} {\bibfield  {journal}
  {\bibinfo  {journal} {J. Phys. Conf. Ser.}\ }\textbf {\bibinfo {volume}
  {2038}},\ \bibinfo {pages} {012010} (\bibinfo {year} {2021})},\ \Eprint
  {http://arxiv.org/abs/2103.13519} {arXiv:2103.13519 [hep-th]} \BibitemShut
  {NoStop}%
\bibitem [{\citenamefont {Moshe}\ and\ \citenamefont
  {Zinn-Justin}(2003)}]{Moshe:2003xn}%
  \BibitemOpen
  \bibfield  {author} {\bibinfo {author} {\bibfnamefont {M.}~\bibnamefont
  {Moshe}}\ and\ \bibinfo {author} {\bibfnamefont {J.}~\bibnamefont
  {Zinn-Justin}},\ }\href {\doibase 10.1016/S0370-1573(03)00263-1} {\bibfield
  {journal} {\bibinfo  {journal} {Phys. Rept.}\ }\textbf {\bibinfo {volume}
  {385}},\ \bibinfo {pages} {69} (\bibinfo {year} {2003})},\ \Eprint
  {http://arxiv.org/abs/hep-th/0306133} {arXiv:hep-th/0306133} \BibitemShut
  {NoStop}%
\bibitem [{\citenamefont {Bonzom}\ \emph {et~al.}(2011)\citenamefont {Bonzom},
  \citenamefont {Gurau}, \citenamefont {Riello},\ and\ \citenamefont
  {Rivasseau}}]{Bonzom:2011zz}%
  \BibitemOpen
  \bibfield  {author} {\bibinfo {author} {\bibfnamefont {V.}~\bibnamefont
  {Bonzom}}, \bibinfo {author} {\bibfnamefont {R.}~\bibnamefont {Gurau}},
  \bibinfo {author} {\bibfnamefont {A.}~\bibnamefont {Riello}}, \ and\ \bibinfo
  {author} {\bibfnamefont {V.}~\bibnamefont {Rivasseau}},\ }\href {\doibase
  10.1016/j.nuclphysb.2011.07.022} {\bibfield  {journal} {\bibinfo  {journal}
  {Nucl. Phys. B}\ }\textbf {\bibinfo {volume} {853}},\ \bibinfo {pages} {174}
  (\bibinfo {year} {2011})},\ \Eprint {http://arxiv.org/abs/1105.3122}
  {arXiv:1105.3122 [hep-th]} \BibitemShut {NoStop}%
\bibitem [{\citenamefont {Gurau}\ and\ \citenamefont
  {Ryan}(2012)}]{Gurau:2011xp}%
  \BibitemOpen
  \bibfield  {author} {\bibinfo {author} {\bibfnamefont {R.}~\bibnamefont
  {Gurau}}\ and\ \bibinfo {author} {\bibfnamefont {J.~P.}\ \bibnamefont
  {Ryan}},\ }\href {\doibase 10.3842/SIGMA.2012.020} {\bibfield  {journal}
  {\bibinfo  {journal} {SIGMA}\ }\textbf {\bibinfo {volume} {8}},\ \bibinfo
  {pages} {020} (\bibinfo {year} {2012})},\ \Eprint
  {http://arxiv.org/abs/1109.4812} {arXiv:1109.4812 [hep-th]} \BibitemShut
  {NoStop}%
\bibitem [{\citenamefont {'t~Hooft}(1974)}]{tHooft:1973alw}%
  \BibitemOpen
  \bibfield  {author} {\bibinfo {author} {\bibfnamefont {G.}~\bibnamefont
  {'t~Hooft}},\ }\href {\doibase 10.1016/0550-3213(74)90154-0} {\bibfield
  {journal} {\bibinfo  {journal} {Nucl. Phys. B}\ }\textbf {\bibinfo {volume}
  {72}},\ \bibinfo {pages} {461} (\bibinfo {year} {1974})}\BibitemShut
  {NoStop}%
\bibitem [{\citenamefont {Brezin}\ \emph {et~al.}(1978)\citenamefont {Brezin},
  \citenamefont {Itzykson}, \citenamefont {Parisi},\ and\ \citenamefont
  {Zuber}}]{Brezin:1977sv}%
  \BibitemOpen
  \bibfield  {author} {\bibinfo {author} {\bibfnamefont {E.}~\bibnamefont
  {Brezin}}, \bibinfo {author} {\bibfnamefont {C.}~\bibnamefont {Itzykson}},
  \bibinfo {author} {\bibfnamefont {G.}~\bibnamefont {Parisi}}, \ and\ \bibinfo
  {author} {\bibfnamefont {J.~B.}\ \bibnamefont {Zuber}},\ }\href {\doibase
  10.1007/BF01614153} {\bibfield  {journal} {\bibinfo  {journal} {Commun. Math.
  Phys.}\ }\textbf {\bibinfo {volume} {59}},\ \bibinfo {pages} {35} (\bibinfo
  {year} {1978})}\BibitemShut {NoStop}%
\bibitem [{\citenamefont {Witten}(2019)}]{Witten:2016iux}%
  \BibitemOpen
  \bibfield  {author} {\bibinfo {author} {\bibfnamefont {E.}~\bibnamefont
  {Witten}},\ }\href {\doibase 10.1088/1751-8121/ab3752} {\bibfield  {journal}
  {\bibinfo  {journal} {J. Phys. A}\ }\textbf {\bibinfo {volume} {52}},\
  \bibinfo {pages} {474002} (\bibinfo {year} {2019})},\ \Eprint
  {http://arxiv.org/abs/1610.09758} {arXiv:1610.09758 [hep-th]} \BibitemShut
  {NoStop}%
\bibitem [{\citenamefont {Sachdev}\ and\ \citenamefont
  {Ye}(1993)}]{Sachdev:1992fk}%
  \BibitemOpen
  \bibfield  {author} {\bibinfo {author} {\bibfnamefont {S.}~\bibnamefont
  {Sachdev}}\ and\ \bibinfo {author} {\bibfnamefont {J.}~\bibnamefont {Ye}},\
  }\href {\doibase 10.1103/PhysRevLett.70.3339} {\bibfield  {journal} {\bibinfo
   {journal} {Phys. Rev. Lett.}\ }\textbf {\bibinfo {volume} {70}},\ \bibinfo
  {pages} {3339} (\bibinfo {year} {1993})},\ \Eprint
  {http://arxiv.org/abs/cond-mat/9212030} {arXiv:cond-mat/9212030} \BibitemShut
  {NoStop}%
\bibitem [{\citenamefont {Kitaev}(2015)}]{Kitaev}%
  \BibitemOpen
  \bibfield  {author} {\bibinfo {author} {\bibfnamefont {A.}~\bibnamefont
  {Kitaev}},\ }\href@noop {} {\bibfield  {journal} {\bibinfo  {journal} {KITP
  strings seminar and Entanglement 2015}\ } (\bibinfo {year} {Feb. 12, April 7,
  and May 27, 2015})}\BibitemShut {NoStop}%
\bibitem [{\citenamefont {Klebanov}\ and\ \citenamefont
  {Tarnopolsky}(2017)}]{Klebanov:2016xxf}%
  \BibitemOpen
  \bibfield  {author} {\bibinfo {author} {\bibfnamefont {I.~R.}\ \bibnamefont
  {Klebanov}}\ and\ \bibinfo {author} {\bibfnamefont {G.}~\bibnamefont
  {Tarnopolsky}},\ }\href {\doibase 10.1103/PhysRevD.95.046004} {\bibfield
  {journal} {\bibinfo  {journal} {Phys. Rev. D}\ }\textbf {\bibinfo {volume}
  {95}},\ \bibinfo {pages} {046004} (\bibinfo {year} {2017})},\ \Eprint
  {http://arxiv.org/abs/1611.08915} {arXiv:1611.08915 [hep-th]} \BibitemShut
  {NoStop}%
\bibitem [{\citenamefont {Giombi}\ \emph {et~al.}(2017)\citenamefont {Giombi},
  \citenamefont {Klebanov},\ and\ \citenamefont
  {Tarnopolsky}}]{Giombi:2017dtl}%
  \BibitemOpen
  \bibfield  {author} {\bibinfo {author} {\bibfnamefont {S.}~\bibnamefont
  {Giombi}}, \bibinfo {author} {\bibfnamefont {I.~R.}\ \bibnamefont
  {Klebanov}}, \ and\ \bibinfo {author} {\bibfnamefont {G.}~\bibnamefont
  {Tarnopolsky}},\ }\href {\doibase 10.1103/PhysRevD.96.106014} {\bibfield
  {journal} {\bibinfo  {journal} {Phys. Rev. D}\ }\textbf {\bibinfo {volume}
  {96}},\ \bibinfo {pages} {106014} (\bibinfo {year} {2017})},\ \Eprint
  {http://arxiv.org/abs/1707.03866} {arXiv:1707.03866 [hep-th]} \BibitemShut
  {NoStop}%
\bibitem [{\citenamefont {Benedetti}\ \emph {et~al.}(2019)\citenamefont
  {Benedetti}, \citenamefont {Gurau},\ and\ \citenamefont
  {Harribey}}]{Benedetti:2019eyl}%
  \BibitemOpen
  \bibfield  {author} {\bibinfo {author} {\bibfnamefont {D.}~\bibnamefont
  {Benedetti}}, \bibinfo {author} {\bibfnamefont {R.}~\bibnamefont {Gurau}}, \
  and\ \bibinfo {author} {\bibfnamefont {S.}~\bibnamefont {Harribey}},\ }\href
  {\doibase 10.1007/JHEP06(2019)053} {\bibfield  {journal} {\bibinfo  {journal}
  {JHEP}\ }\textbf {\bibinfo {volume} {06}},\ \bibinfo {pages} {053} (\bibinfo
  {year} {2019})},\ \Eprint {http://arxiv.org/abs/1903.03578} {arXiv:1903.03578
  [hep-th]} \BibitemShut {NoStop}%
\bibitem [{\citenamefont {Benedetti}\ \emph {et~al.}(2022)\citenamefont
  {Benedetti}, \citenamefont {Gurau}, \citenamefont {Harribey},\ and\
  \citenamefont {Lettera}}]{Benedetti:2021wzt}%
  \BibitemOpen
  \bibfield  {author} {\bibinfo {author} {\bibfnamefont {D.}~\bibnamefont
  {Benedetti}}, \bibinfo {author} {\bibfnamefont {R.}~\bibnamefont {Gurau}},
  \bibinfo {author} {\bibfnamefont {S.}~\bibnamefont {Harribey}}, \ and\
  \bibinfo {author} {\bibfnamefont {D.}~\bibnamefont {Lettera}},\ }\href
  {\doibase 10.1007/JHEP02(2022)147} {\bibfield  {journal} {\bibinfo  {journal}
  {JHEP}\ }\textbf {\bibinfo {volume} {02}},\ \bibinfo {pages} {147} (\bibinfo
  {year} {2022})},\ \Eprint {http://arxiv.org/abs/2111.11792} {arXiv:2111.11792
  [hep-th]} \BibitemShut {NoStop}%
\bibitem [{\citenamefont {de~Forcrand}(2009)}]{deForcrand:2009zkb}%
  \BibitemOpen
  \bibfield  {author} {\bibinfo {author} {\bibfnamefont {P.}~\bibnamefont
  {de~Forcrand}},\ }\href {\doibase 10.22323/1.091.0010} {\bibfield  {journal}
  {\bibinfo  {journal} {PoS}\ }\textbf {\bibinfo {volume} {LAT2009}},\ \bibinfo
  {pages} {010} (\bibinfo {year} {2009})},\ \Eprint
  {http://arxiv.org/abs/1005.0539} {arXiv:1005.0539 [hep-lat]} \BibitemShut
  {NoStop}%
\bibitem [{\citenamefont {Cornwall}\ \emph {et~al.}(1974)\citenamefont
  {Cornwall}, \citenamefont {Jackiw},\ and\ \citenamefont
  {Tomboulis}}]{Cornwall:1974vz}%
  \BibitemOpen
  \bibfield  {author} {\bibinfo {author} {\bibfnamefont {J.~M.}\ \bibnamefont
  {Cornwall}}, \bibinfo {author} {\bibfnamefont {R.}~\bibnamefont {Jackiw}}, \
  and\ \bibinfo {author} {\bibfnamefont {E.}~\bibnamefont {Tomboulis}},\ }\href
  {\doibase 10.1103/PhysRevD.10.2428} {\bibfield  {journal} {\bibinfo
  {journal} {Phys. Rev. D}\ }\textbf {\bibinfo {volume} {10}},\ \bibinfo
  {pages} {2428} (\bibinfo {year} {1974})}\BibitemShut {NoStop}%
\bibitem [{\citenamefont {Benedetti}\ and\ \citenamefont
  {Gurau}(2018)}]{Benedetti:2018goh}%
  \BibitemOpen
  \bibfield  {author} {\bibinfo {author} {\bibfnamefont {D.}~\bibnamefont
  {Benedetti}}\ and\ \bibinfo {author} {\bibfnamefont {R.}~\bibnamefont
  {Gurau}},\ }\href {\doibase 10.1007/JHEP05(2018)156} {\bibfield  {journal}
  {\bibinfo  {journal} {JHEP}\ }\textbf {\bibinfo {volume} {05}},\ \bibinfo
  {pages} {156} (\bibinfo {year} {2018})},\ \Eprint
  {http://arxiv.org/abs/1802.05500} {arXiv:1802.05500 [hep-th]} \BibitemShut
  {NoStop}%
\bibitem [{\citenamefont {Berges}(2004)}]{Berges:2004yj}%
  \BibitemOpen
  \bibfield  {author} {\bibinfo {author} {\bibfnamefont {J.}~\bibnamefont
  {Berges}},\ }\href {\doibase 10.1063/1.1843591} {\bibfield  {journal}
  {\bibinfo  {journal} {AIP Conf. Proc.}\ }\textbf {\bibinfo {volume} {739}},\
  \bibinfo {pages} {3} (\bibinfo {year} {2004})},\ \Eprint
  {http://arxiv.org/abs/hep-ph/0409233} {arXiv:hep-ph/0409233} \BibitemShut
  {NoStop}%
\bibitem [{\citenamefont {Benedetti}\ \emph {et~al.}(2021)\citenamefont
  {Benedetti}, \citenamefont {Gurau},\ and\ \citenamefont
  {Harribey}}]{Benedetti:2020sye}%
  \BibitemOpen
  \bibfield  {author} {\bibinfo {author} {\bibfnamefont {D.}~\bibnamefont
  {Benedetti}}, \bibinfo {author} {\bibfnamefont {R.}~\bibnamefont {Gurau}}, \
  and\ \bibinfo {author} {\bibfnamefont {S.}~\bibnamefont {Harribey}},\ }\href
  {\doibase 10.1103/PhysRevD.103.046018} {\bibfield  {journal} {\bibinfo
  {journal} {Phys. Rev. D}\ }\textbf {\bibinfo {volume} {103}},\ \bibinfo
  {pages} {046018} (\bibinfo {year} {2021})},\ \Eprint
  {http://arxiv.org/abs/2011.11276} {arXiv:2011.11276 [hep-th]} \BibitemShut
  {NoStop}%
\bibitem [{\citenamefont {Christiansen}\ \emph {et~al.}(2018)\citenamefont
  {Christiansen}, \citenamefont {Litim}, \citenamefont {Pawlowski},\ and\
  \citenamefont {Reichert}}]{Christiansen:2017cxa}%
  \BibitemOpen
  \bibfield  {author} {\bibinfo {author} {\bibfnamefont {N.}~\bibnamefont
  {Christiansen}}, \bibinfo {author} {\bibfnamefont {D.~F.}\ \bibnamefont
  {Litim}}, \bibinfo {author} {\bibfnamefont {J.~M.}\ \bibnamefont
  {Pawlowski}}, \ and\ \bibinfo {author} {\bibfnamefont {M.}~\bibnamefont
  {Reichert}},\ }\href {\doibase 10.1103/PhysRevD.97.106012} {\bibfield
  {journal} {\bibinfo  {journal} {Phys. Rev. D}\ }\textbf {\bibinfo {volume}
  {97}},\ \bibinfo {pages} {106012} (\bibinfo {year} {2018})},\ \Eprint
  {http://arxiv.org/abs/1710.04669} {arXiv:1710.04669 [hep-th]} \BibitemShut
  {NoStop}%
\bibitem [{\citenamefont {Zinn-Justin}(2021)}]{Zinn-Justin:1989rgp}%
  \BibitemOpen
  \bibfield  {author} {\bibinfo {author} {\bibfnamefont {J.}~\bibnamefont
  {Zinn-Justin}},\ }\href@noop {} {\emph {\bibinfo {title} {{Quantum field
  theory and critical phenomena}}}},\ \bibinfo {series} {International Series
  of Monographs on Physics}, Vol.~\bibinfo {volume} {77}\ (\bibinfo
  {publisher} {Oxford University Press},\ \bibinfo {year} {2021})\BibitemShut
  {NoStop}%
\bibitem [{\citenamefont {Fisher}(1978)}]{Fisher:1978pf}%
  \BibitemOpen
  \bibfield  {author} {\bibinfo {author} {\bibfnamefont {M.~E.}\ \bibnamefont
  {Fisher}},\ }\href {\doibase 10.1103/PhysRevLett.40.1610} {\bibfield
  {journal} {\bibinfo  {journal} {Phys. Rev. Lett.}\ }\textbf {\bibinfo
  {volume} {40}},\ \bibinfo {pages} {1610} (\bibinfo {year}
  {1978})}\BibitemShut {NoStop}%
\bibitem [{\citenamefont {Gromov}\ \emph {et~al.}(2019)\citenamefont {Gromov},
  \citenamefont {Kazakov},\ and\ \citenamefont {Korchemsky}}]{Gromov:2018hut}%
  \BibitemOpen
  \bibfield  {author} {\bibinfo {author} {\bibfnamefont {N.}~\bibnamefont
  {Gromov}}, \bibinfo {author} {\bibfnamefont {V.}~\bibnamefont {Kazakov}}, \
  and\ \bibinfo {author} {\bibfnamefont {G.}~\bibnamefont {Korchemsky}},\
  }\href {\doibase 10.1007/JHEP08(2019)123} {\bibfield  {journal} {\bibinfo
  {journal} {JHEP}\ }\textbf {\bibinfo {volume} {08}},\ \bibinfo {pages} {123}
  (\bibinfo {year} {2019})},\ \Eprint {http://arxiv.org/abs/1808.02688}
  {arXiv:1808.02688 [hep-th]} \BibitemShut {NoStop}%
\bibitem [{\citenamefont {Casalderrey-Solana}\ \emph
  {et~al.}(2014)\citenamefont {Casalderrey-Solana}, \citenamefont {Liu},
  \citenamefont {Mateos}, \citenamefont {Rajagopal},\ and\ \citenamefont
  {Wiedemann}}]{Casalderrey-Solana:2011dxg}%
  \BibitemOpen
  \bibfield  {author} {\bibinfo {author} {\bibfnamefont {J.}~\bibnamefont
  {Casalderrey-Solana}}, \bibinfo {author} {\bibfnamefont {H.}~\bibnamefont
  {Liu}}, \bibinfo {author} {\bibfnamefont {D.}~\bibnamefont {Mateos}},
  \bibinfo {author} {\bibfnamefont {K.}~\bibnamefont {Rajagopal}}, \ and\
  \bibinfo {author} {\bibfnamefont {U.~A.}\ \bibnamefont {Wiedemann}},\ }\href
  {\doibase 10.1017/CBO9781139136747} {\emph {\bibinfo {title} {{Gauge/String
  Duality, Hot QCD and Heavy Ion Collisions}}}}\ (\bibinfo  {publisher}
  {Cambridge University Press},\ \bibinfo {year} {2014})\ \Eprint
  {http://arxiv.org/abs/1101.0618} {arXiv:1101.0618 [hep-th]} \BibitemShut
  {NoStop}%
\bibitem [{\citenamefont {Berges}\ \emph {et~al.}(2021)\citenamefont {Berges},
  \citenamefont {Heller}, \citenamefont {Mazeliauskas},\ and\ \citenamefont
  {Venugopalan}}]{Berges:2020fwq}%
  \BibitemOpen
  \bibfield  {author} {\bibinfo {author} {\bibfnamefont {J.}~\bibnamefont
  {Berges}}, \bibinfo {author} {\bibfnamefont {M.~P.}\ \bibnamefont {Heller}},
  \bibinfo {author} {\bibfnamefont {A.}~\bibnamefont {Mazeliauskas}}, \ and\
  \bibinfo {author} {\bibfnamefont {R.}~\bibnamefont {Venugopalan}},\ }\href
  {\doibase 10.1103/RevModPhys.93.035003} {\bibfield  {journal} {\bibinfo
  {journal} {Rev. Mod. Phys.}\ }\textbf {\bibinfo {volume} {93}},\ \bibinfo
  {pages} {035003} (\bibinfo {year} {2021})},\ \Eprint
  {http://arxiv.org/abs/2005.12299} {arXiv:2005.12299 [hep-th]} \BibitemShut
  {NoStop}%
\end{thebibliography}%
	
\end{document}